# GRAVITATIONAL LENSES AND THE STRUCTURE OF GALAXIES

CHRISTOPHER S. KOCHANEK

*Harvard-Smithsonian Center for Astrophysics*
*60 Garden Street, Cambridge, MA 02138, USA*

**Abstract.** Nearly singular isothermal mass distributions with small core radii are consistent with stellar dynamics, lens statistics, and lens models as a model for E/S0 galaxies. Models like the de Vaucouleurs model with a constant mass-to-light ratio are not. While the isothermal distributions are probably an oversimplification, E/S0 galaxies (at least in projection) must have significant amounts of dark matter on scales of an effective radius.

## 1. Inferences From Dynamics

The mass distribution in E/S0 galaxies remains unclear despite many years of effort using stellar dynamics and other tracers. It is probable from X-ray studies and the rare polar ring galaxies (see Sackett in this volume) that the outer regions are dominated by dark matter, while the inner regions are consistent with either constant mass-to-light ratio ($M/L$) models or dark matter models. In addition to any intrinsic interest in the matter distribution of E/S0 galaxies, their structure is of crucial importance in using gravitational lens statistics to determine the cosmological model.

The state of the art, constant mass-to-light ratio, dynamical model for E/S0 galaxies that is fit to observational data is the two-integral axisymmetric model. In a survey of some forty galaxies, van der Marel (1991) derived a mass-to-light ratio of $(M/L)_B = (10 \pm 2)h$ for an $L_*$ galaxy using these dynamical models. The more traditional lensing model is the singular isothermal sphere (SIS), and models of E/S0 galaxies with this mass distribution find that the velocity dispersion of the dark matter is $\sigma_{DM*} \simeq 225 \pm 20$ km s$^{-1}$ (Kochanek 1994, Breimer & Sanders 1993, Franx 1993), which is approximately equal to the central velocity dispersion of the stars, $\sigma_c$. Earlier models with $\sigma_{DM*} = (3/2)^{1/2}\sigma_c$ (e.g. Turner et al. 1984) are based on oversimplified dynamical models (see Kochanek in this volume). Normal dynamical techniques have difficulty determining which of these two extreme models actually applies to E/S0 galaxies.

## 2. Inferences From Lens Statistics

Maoz & Rix (1993) and Kochanek (1993, 1995b), made detailed statistical models of the observed lens samples for various mass distributions (also see Rix, Kochanek, and Claeskens et al. in this volume). These studies examined a range of models from de Vaucouleurs models to softened isothermal spheres, emphasizing the limits on the cosmological constant. The cosmological conclusions are independent of the mass distribution used for the lens galaxies.

The normalization of the galaxy masses is determined by the distribution of image separations found in the surveys. Moaz & Rix (1993) first pointed out that de Vaucouleurs models normalized by the mass-to-light ratios estimated from dynamical models of nearby E/S0 galaxies (e.g. van der Marel 1991) produced image separations that were too small to fit the lens data. Kochanek (1995b) demonstrated that a mass-to-light ratio of $(M/L)_B \simeq (25 \pm 5)h$ at 90% confidence is required to fit the separations, compared to $(10 \pm 2)h$ in the dynamical models. For the softened isothermal sphere (Kochanek 1993, 1995b), fitting the separations requires $\sigma_{DM*} = 220 \pm 20$ km s$^{-1}$, consistent with the dynamical models.[1] The isothermal lens must be nearly singular to avoid the appearance of central images in the lenses (e.g., Wallington & Narayan 1993, Kassiola & Kovner 1993).

At least in theory, large amounts of evolution (Mao & Kochanek 1993, Rix et al. 1994), extreme errors in the selection function, or (for optical lenses) extinction (Tomita 1995, Kochanek 1995b) can invalidate the statistical inferences. In practice, however, the are sufficient constraints in the current data to rule out any dramatic errors. Unfortunately, the statistical models of the lens surveys cannot as yet differentiate between mass distributions except by comparing the normalization required by the lens data to the normalization required by stellar dynamics. Hopefully, models of individual lens systems can both validate the normalization and differentiate between radial mass distributions.

---

[1] Maoz & Rix (1993) added a softened isothermal halo to the de Vaucouleurs models, but the resulting deflection produced by the lens so closely resembles that of a softened isothermal sphere that there is no point in treating them as a separate class of models.

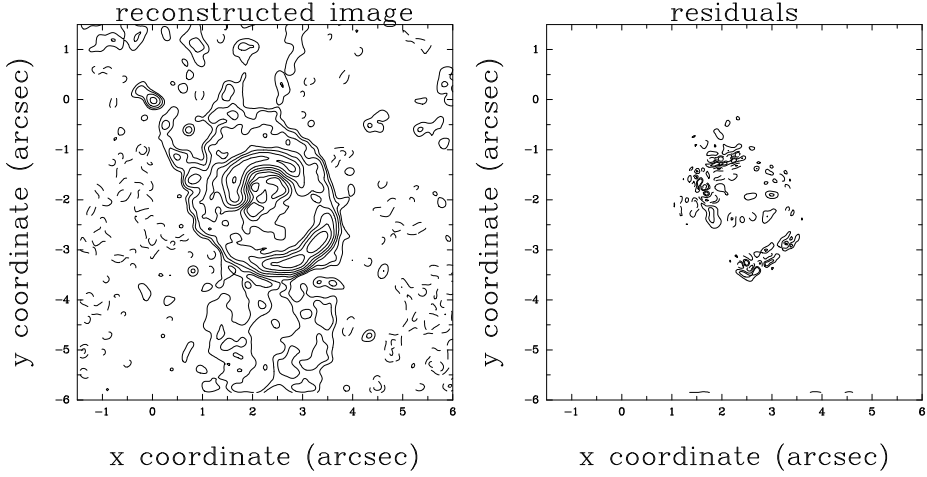

*Figure 1.* Panel (a) shows the reconstructed image for the best fit expanded ellipsoid model, and Panel (b) shows the residuals. The largest positive and negative residuals are 139 $\mu$Jy/pixel and $-191$ $\mu$Jy/pixel. The dashed contours are drawn at $-70$ and $-35$ $\mu$Jy/pixel and the solid contours are drawn at 35, 70, 140, 280, 560, 1120, 2240, and 4480 $\mu$Jy/pixel. The estimated noise in the map is 35 $\mu$Jy/pixel, so the contours lie at $\pm 1$, $\pm 2$, 4, 8, 16, 32, 64, and 128 times the estimated noise in the map.

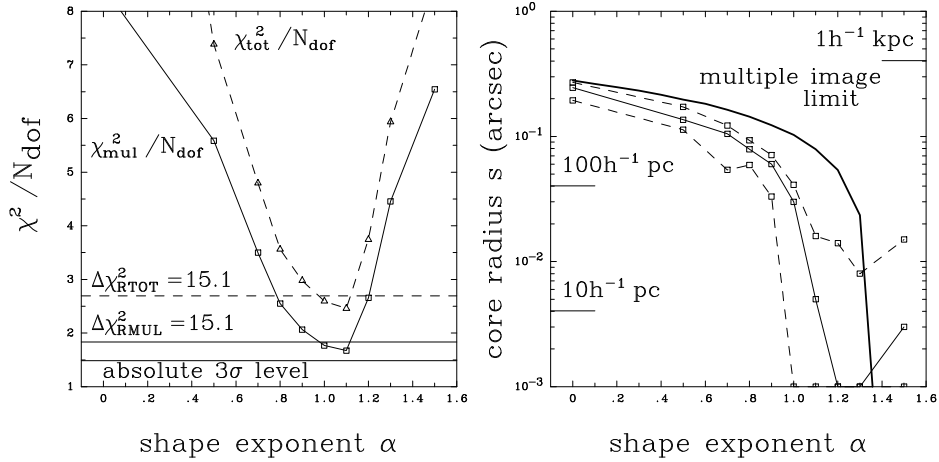

*Figure 2.* Panel (a) shows the minimum values of $\chi^2_{mul}/N_{dof}$ (solid/points) and $\chi^2_{tot}/N_{dof}$ (dashed/points) as a function of the exponent $\alpha$. The core radius of each model has been optimized. The bottom horizontal line shows the formal $3\sigma$ deviation of $\chi^2/N_{dof}$ from unity. The other two horizontal lines show where $\Delta\chi^2_r = 15.1$ (formally a 99.99% change). Panel (b) shows the optimized value for the core radius (solid/squares) and the limits (dashed/squares) on the core radius for $\Delta\chi^2_r = 15.1$ in the rescaled $\chi^2_r$ estimator. The heavy solid line shows the upper limit on the core radius if the multiply imaged region is larger than 1.5 arcseconds.

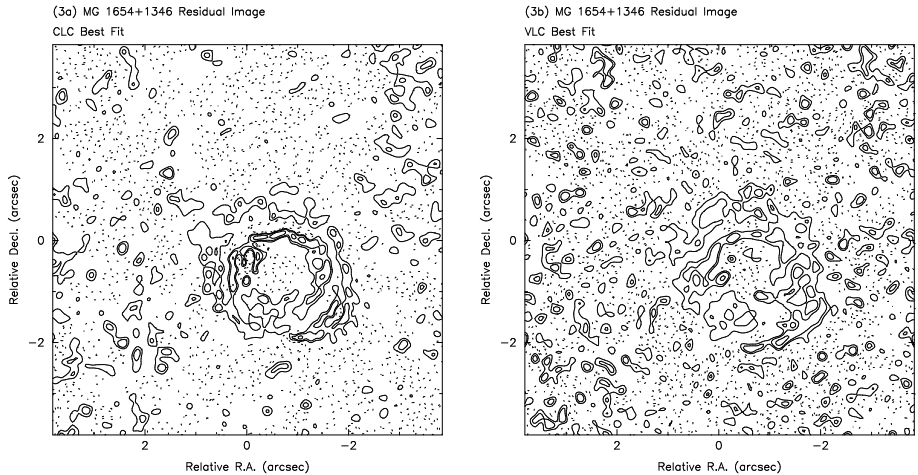

*Figure 3.* A comparison of the residuals in fitting MG 1654+134 using the Clean map (a) or the raw visibilities/"dirty map" (b). The contour levels are drawn at $\pm 1$, $\pm 2$, 4, 8, 16, 32, 64 and 96 percent of the peak in the Clean map. The estimated noise is 59 $\mu$Jy/Beam. In (a) the minimum, maximum, and rms errors are -411 $\mu$Jy/Beam, 315 $\mu$Jy/Beam, and 71 $\mu$Jy/Beam, while in (b) they are -211 $\mu$Jy/Beam, 247 $\mu$Jy/Beam, and 61 $\mu$Jy/Beam.

## 3. Inferences from Lens Models

Not all lenses are useful for distinguishing between radial mass distributions. Point lenses without VLBI transformation matrices are particularly ill-suited to this problem because the fit is either under-constrained (2 image systems) or dominated by the details of the angular structure (4 image systems). The utility of VLBI transformation matrices in avoiding this problem has yet to be seriously investigated. Some of the radio rings, however, have extended emission spread over a wide range of radii from the centers of the lenses. These are the most promising systems for studying the radial mass distribution. Unfortunately, modeling such systems given the finite instrumental resolution is complicated (see Kochanek & Narayan 1992, Wallington et al. 1994, 1995, and Ellithorpe et al. 1995).

MG 1654+134 was found in the MG survey (Langston et al. 1989, 1990). One lobe of a double lobed, $z = 1.74$ radio quasar is lensed into a 2 arcsec diameter ring around a r=18.7 mag, $z = 0.254$ galaxy. The emission in the ring is very extended, making it an ideal candidate for estimating the radial mass distribution of the lens. Kochanek (1995a) treated two general classes of models for the radial profile: the de Vaucouleurs model and softened power-law density distributions of the form $\Sigma \propto (r^2 + s^2)^{\alpha/2-1}$, where the isothermal distribution has $\alpha = 1$.

The first question we examine is the normalization of the models or the mass inside the ring. We find that the mass inside radius $r = 0.9$ arcsec from the lens galaxy is $M = (7.75 \pm 0.03)h^{-1}10^{10}M_\odot$ at 90% confidence for $\Omega = 1$. The systematic uncertainty from the cosmological model for the range $0 < \Omega < 1$ is 7%. This corresponds to a blue mass-to-light ratio inside the ring of $(M/L)_B = (20.4 \pm 2.8)(f_e/1.4)h$ where the uncertainty is entirely due to the uncertainties in the enclosed light (also see Burke et al. 1992). Equivalently, the velocity dispersion of an isothermal model must be $\sigma_{DM} = (223 \pm 11)(f_e/1.4)^{-0.28}$ km s$^{-1}$. The factor $f_e$ corrects for the expected fading of a passively evolving elliptical between the lens redshift and the current epoch. Both of these measurements match the results found in the statistical studies.

Figure 1 shows the best fit model with $\alpha = 1$ for MG1654+134 and its residuals. To the eye, the reconstructed image is indistinguishable from the original images (despite the logarithmically spaced contours), but there is a pattern of residuals surrounding the ring at the level of a few standard deviations above the noise. Figure 2 shows the $\chi^2$ of the fit as a function of $\alpha$. The allowed models have $\alpha \simeq 1.0 \pm 0.1$, corresponding to an isothermal distribution. Figure 2 also shows the best fit core radius $s$ and its error bars as a function of the exponent $\alpha$. The core radius must be very compact for isothermal lenses, with $s \lesssim 0.02$ arcsec or approximately $50h^{-1}$ pc, but it must have a large, finite value for the more centrally concentrated models. The best fit de Vaucouleurs model has a $\chi^2$ slightly worse than the models in the permitted range for $\alpha$. Inside the ring, this model closely matches the deflection profile of the isothermal model, but it drops too rapidly outside the ring. This is the same problem that makes the de Vaucouleurs models incapable of fitting the distribution of lens separations in the statistical studies.

We have repeated the calculations for the nearly isothermal lenses using a variant that fits the raw measured visibilities rather than the processed Clean map (the VLC algorithm, Ellithorpe et al. 1995) and a variant using the maximum entropy method (the LensMEM algorithm, Wallington et al. 1994, 1995) with the same results. While the parameters of the best fit models for each algorithm are mutually consistent, the residuals in the VLC inversions are significantly lower than in the inversions starting from the processed Clean map (see Figure 3). This means that the Clean process can introduce significant and detectable artifacts into maps of lensed images, and accurate models must start from the visibility data. Further experiments with self-calibration showed evidence that the self-calibration step also introduces detectable artifacts (Ellithorpe et al. 1995). A map of a lens produced with a good, if approximate, lens model is generally a better approximation to the true image than a simple Clean map. Even an approximate lens model begins to enforce the constraints required of a real lensed image.

## 4. Conclusions

As far as we can tell, the only model that is currently consistent with stellar dynamics, lens statistics, and lens models is a mass distribution similar to the nearly singular isothermal sphere. This is, undoubtedly, an oversimplification of the true mass distribution. All is not perfect, however. For example, we have no good model for the ring lens MG1131+0456 (Chen et al. 1995), and there is some evidence that the ellipticities required to fit the lenses are higher than is reasonable for the observed ellipticities of the luminous material. Some of this is due to the oversimplifications of the elliptical structures of the models, but it deserves further attention. A very interesting study that has yet to be done is to obtain data on the velocity dispersion profiles in the ring lenses and directly compare the inferences from stellar dynamics and lens models.